\documentclass{ws-procs975x65}

\begin{document}

\title{HIERARCHICAL CLUSTERING AND THE BAO SIGNATURE}

\author{WOJCIECH A. HELLWING$^*$}

\address{
Interdisciplinary Centre for Mathematical and Computational Modeling (ICM),\\
University of Warsaw, ul. Pawi\'nskiego 5a, Warsaw, Poland,$^*$E-mail: pchela@icm.edu.pl
}

\author{ROMAN JUSZKIEWICZ}

\address{
Institute of Astronomy,University of Zielona G\'ora,\\
ul. Lubuska 2, Zielona G\'ora, Poland\\
}

\author{RIEN VAN DE WEYGAERT}

\address{
Kapteyn Astronomical Institute, University of Groningen,\\
P.O. Box 800, 9725LB Groningen, the Netherlands\\
}

\author{MACIEJ BILICKI}

\address{
Astrophysics, Cosmology and Gravity Centre (ACGC),Department of Astronomy,\\
University of Cape Town,Private Bag X3, Rondebosch 7701, South Africa
}

\begin{abstract}
In this contribution we present the preliminary results regarding
the non-linear BAO signal in higher-order statistics of the
cosmic density field. We use ensembles of N-body simulations to show
that the non-linear evolution changes the amplitudes of the BAO
signal, but has a negligible effect on the scale of the BAO feature.
The latter observation accompanied by the fact that the BAO feature
amplitude roughly doubles as one moves to higher orders, suggests
that the higher-order correlation amplitudes can be used as probe
  of the BAO signal.
\end{abstract}


\bodymatter

\section{Introduction}
\label{sec1}
Thanks to the impressive development in observational astronomy
that we have witnessed in the last two decades, we are living now
in the era of the precision cosmology. Growing data sets and galaxy 
catalogues allow for extraction and measurement of percent-level signals.
However, this advance of observational data has however not 
always matched by an accompanying increase in theoretical understanding.
Observations of the evolution
and the structure of the large-scale clustering pattern of galaxies 
has left us with the riddle about the nature of the seemingly dominant 
yet elusive components of cosmic energy: dark matter(DM) and dark energy(DE).
One of the main precision probes of 21st century cosmology is that of 
the {\it Baryon Acoustic Oscillations} (BAO).

The BAO feature in the cosmic density field is closely related
to the size of the sound horizon during at recombination
era. The measurements of the size of the BAO feature in the low-redshift 
Universe constrain the cosmic expansion history. Such measurements,
in principle, should have discriminatory power to distinguish between
dynamical DE models and the cosmological constant. 

Measurements of the BAO feature have been reported for the power spectrum
and the two-point correlation function of galaxies \cite{BAOdetect,BAOobs1}.
However there are well-known systematics and non-linear effects that lower
the accuracy of the two-point statistics. The most important among them
are the biasing of galaxies with respect to the dark matter distribution
and redshift space distortions. Recently, we have proposed 
that higher-order statistics of clustering could be a useful
probe of the BAO signal\cite{JHvW} (hereafter JHvW). 
Higher-order clustering amplitudes have the advantage that they 
are less prone to systematics and non-linear effects. On the other hand, 
in general they have a lower signal-to-noise level. 
Here we explore the original idea of JHvW using the N-body 
simulations. Also, we investigate the idea for higher order statistics
beyond that of the three-point clustering measure.

\section{Hierarchical amplitudes and the BAO}
\label{sec2}
We define the volume-averaged $J$-point correlation function (of the density/galaxy field) as 
\begin{equation}
\label{volume-aver-nfunction}
\bar{\xi_J}\,=\,V_W^{-J}\int_S d\mathbf{x_1}...d\mathbf{x_J}W(\mathbf{x_1})...W(\mathbf{x_J})\xi_J(\mathbf{x_1},...,\mathbf{x_J})\,,
\end{equation}
where $\mathbf{x_i}$ is the comoving separation vector, $W(\mathbf{x})$ is a window function with volume $V_W$,
and the integral covers the entire volume $S$. The hierarchical amplitudes $S_n(R)$ of order $n$ are conventionally defined as,
\begin{equation}
\label{eqn:corr_amp}
S_n(R)\,=\,{\displaystyle \bar{\xi_n}\over \displaystyle \bar{\xi_2}^{n-1}}\,=\,\bar{\xi_n}\,\sigma^{-2(n-1)}\;
\end{equation}
with volume-averaged correlation functions $\bar{\xi_n}(R)$ and the variance $\sigma^2(R)$ implicitly depending 
on the smoothing scale $R$. Using perturbation theory (PT), Juszkiewicz {\it et al.}\cite{Juszkiewicz1993} showed that 
the reduced skewness ($S_3$) of the density field is a function of the logarithmic slope of the variance (power spectrum)
at a given smoothing scale $R$. Furthermore, Bernardeau \cite{Bernardeau1994} showed that also higher-order 
clustering amplitudes are functions of higher-order
derivatives of the variance. JHvW noticed that the unique feature of the BAO signal, the so-called {\it wiggles}, should 
be imprinted also in the slope of the variance, and hence in the high-order hierarchical amplitudes.

\section{Results}
\label{sec3}
\begin{figure*}[t]
\begin{center}
\psfig{file=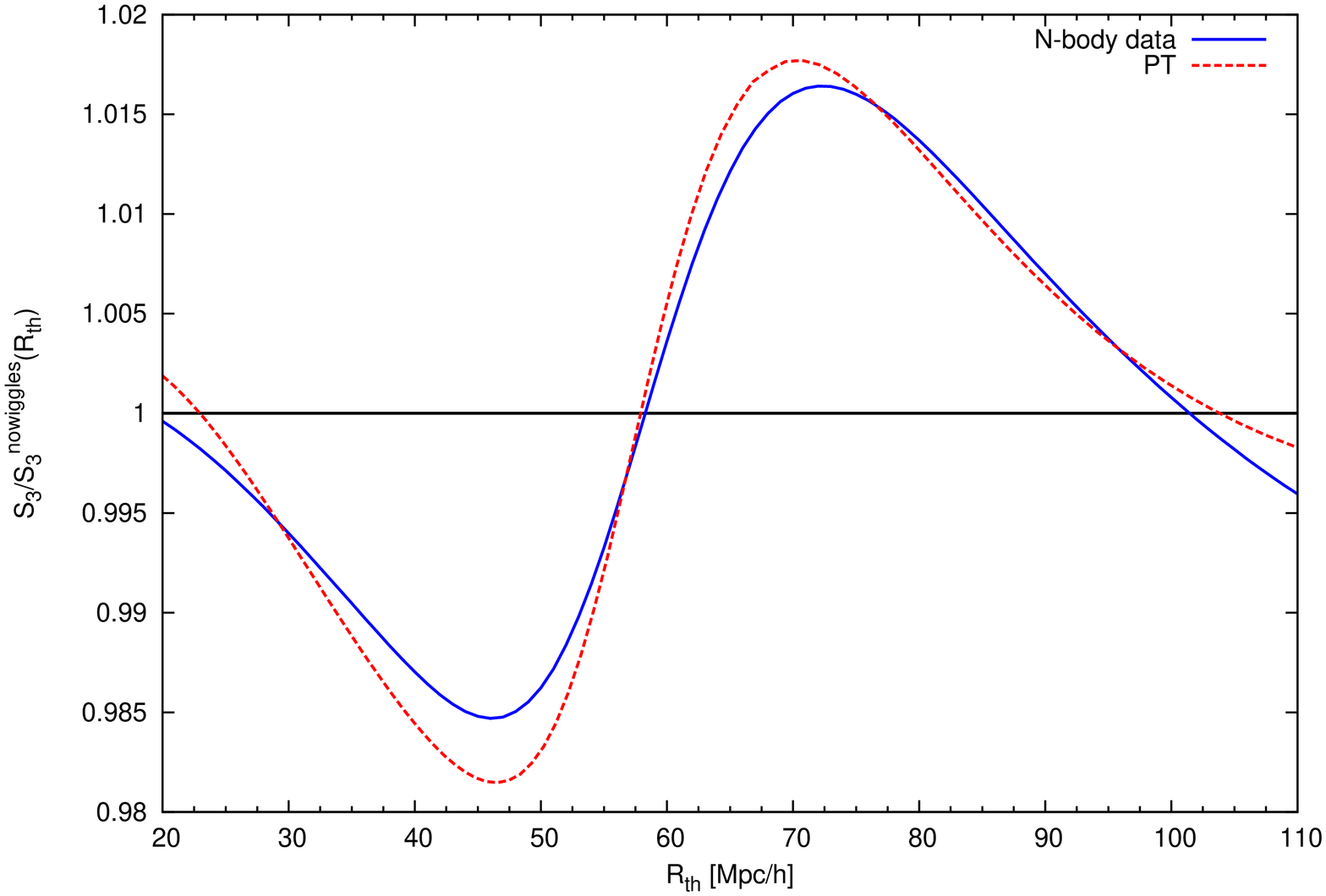,angle=-90,width=2.3in}
\psfig{file=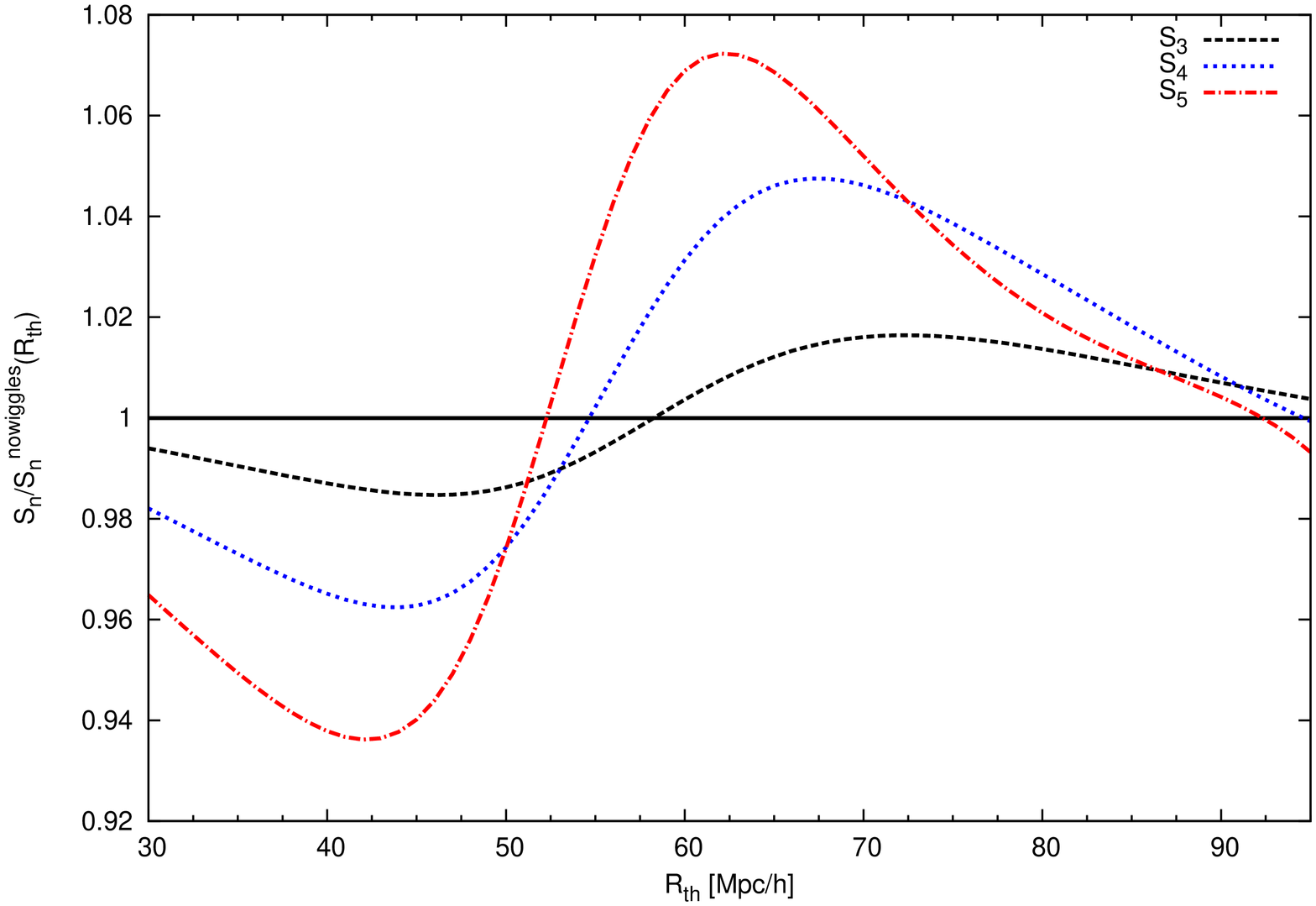,angle=-90,width=2.3in}
\end{center}
\caption{{\it Left}: Comparison of the PT prediction and the fully non-linear evolution for the case of $S_3$, 
{\it Right}: the shape and the amplitude of the BAO signal for $S_3, S_4$ and $S_5$.}
\label{fig1}
\end{figure*}
Here we present the preliminary results of our N-body simulations. We have conducted
high-resolution N-body simulations using the publicly available code by 
Volker Springel - \verb#Gadget2#\cite{Gadget2}.
To assess the exact features resulting from the presence of the BAO signal we have used 
two ensembles of simulations. In one of these, we applied initial conditions generated with 
BAO wiggles smoothed-out (dubbed 'no-wiggles'). In the other,
we implemented the initial density field with fully consistent BAO features. 

Our results are summarised in Figure \ref{fig1}. In the left panel we plot the prediction 
of the BAO signal for the reduced skewness $S_3$ computed using PT, together with the signal
obtained from the simulations. The fully non-linear evolution of the structure formation lowers the 
amplitude of the BAO signal by $\sim\!33\%$ compared to PT. However, an important thing to 
note here is that the non-linear effects seem to have negligible impact on the {\bf scale}
of the BAO feature in the hierarchical amplitude. This has profound consequences, as this is the scale
associated with the BAO signature. We should remind that it is the BAO scale that constrains 
the expansion history, not the amplitude. In the right hand side panel we plot the BAO feature
measured from simulations for the first three amplitudes - $S_3$ (skewness), $S_4$ (kurtosis)
and the $S_5$. The plot shows that when one moves to statistics of one order higher, the amplitude of the BAO signal
doubles. Such a behaviour of the BAO signature in the high-order clustering amplitudes
opens a possibility to extract this signal from the current and future large and deep galaxy redshift surveys,
avoiding to some extent the systematic effects that lower the accuracy of the two-point statistics.

\section*{Acknowledgements}
Soon after we started working on this project, Roman Juszkiewicz had undergone a dramatic deterioration 
of health, as a result of which he passed away on 28th January 2012.  We leave this contribution as 
another tribute to a leading scientist and teacher who was a great friend to the entire community.
\bibliographystyle{ws-procs975x65}
\bibliography{main}

\end{document}